\documentclass[12pt]{JHEP3}
\usepackage{epsfig}
\usepackage{amsmath}
\newcommand{\p}{\partial}

\newcommand{\nn}{\nonumber}

\newcommand{\no}{{}^\circ \hspace{-0.16cm} {}_{\circ}\:}
\title{Brane Decay and Death of Open Strings} 
  
\author{
Koji Hashimoto$^a$ and Seiji Terashima$^b$\\ 
$^a$Institute of Physics, University of Tokyo\\
Komaba, Tokyo 153-8902, Japan\\
E-mail: \email{koji@hep1.c.u-tokyo.ac.jp}\\
$^b$New High Energy Theory Center, Rutgers University\\
126 Frelinghuysen Road, Piscataway, NJ 08854-8019, USA\\
E-mail: \email{seijit@physics.rutgers.edu}\\
}

\abstract{
We show how open strings cease to propagate when unstable D-branes
decay. The information on the propagation is encoded in BSFT two-point
functions for arbitrary profiles of open string excitations. We evaluate
them in tachyon condensation backgrounds corresponding to (i) static
spatial tachyon kink (= lower dimensional BPS D-brane) and (ii)
homogeneous rolling tachyon. For (i) the propagation is restricted to
the directions along the tachyon kink, while for (ii) all the open
string excitations cease to propagate at late time and are subject to a
collapsed light cone characterized by Carrollian contraction of Lorentz
group.} 
 

\preprint{
{\normalsize{\tt hep-th/0404237}}\\ 
{\normalsize UT-Komaba/04-6}
}

\begin{document}

\section{Open strings in tachyon condensation}

What happens to open strings when D-branes decay by tachyon 
condensation?  Although much interesting structures of off-shell string
theory has been revealed since Sen's conjectures \cite{Senconje} on
disappearance of unstable D-branes by open string tachyon condensation, 
the above question which is indispensable for truly understanding the
decay of D-branes is still unanswered in a satisfactory manner. At
vanishing string coupling, the D-branes are by themselves defined as
hypersurfaces on which open strings end, thus the question looks
tautological and ill-defined --- however, this should be the place where
intriguing physics is hidden behind. In spite of the fact that the
degrees of freedom of dynamical D-branes are defined through the open
strings, the theory should also describe physics without D-branes. 
In this article, we provide a description of open strings in brane
decay backgrounds. 

The tachyon condensation is involved intrinsically with off-shell
physics in string theory such as its vacuum structure, one has to
employ string field theories to make an advance to answer the above
question. The string field theories known so far consist mainly of two
categories  --- boundary string field theory (BSFT) \cite{BSFT} and
cubic string field theory (CSFT) \cite{Wittencubic}. Depending on which
one uses, the approaches to the above question might be different. Here
we take the first one since it enables us to observe a direct relation
to worldsheet properties, as we shall see in this article. In the
approaches using the BSFT, the essential point was provided in
\cite{Sen-nonBPS},\footnote{See also \cite{KMM,others, Yi-con, GHoriY}. 
The observation in \cite{Sen-nonBPS} is purely classical, while the
confinement mechanism proposed earlier by Yi \cite{Yi-con} is a quantum
effect.} in 
which the Sen's conjecture on the disappearance of unstable D-branes by
the tachyon condensation \cite{Senconje} is realized in such a way that
the D-brane worldvolume action (BSFT action) itself vanishes. This is a
general property of BSFT actions concerning the tachyon potential in
string theory. When a constant tachyon (which is off-shell) is put to
its vacuum $|T|=\infty$, the BSFT action vanishes due to the overall
potential factor going like  $\sim e^{-T^2}$ or $\sim e^{-T\bar{T}}$ in 
superstring theory. 

The achievements in the BSFT is not only for the constant
tachyon. Tachyon profiles linear in the target space coordinates provide
intriguing results, including a verification of the Sen's conjecture on
the D-brane descent relations \cite{Senconje}, for example. For a
non-BPS D9-brane in type IIA superstring theory, a linear tachyon
profile
\begin{eqnarray}
 T = u_9 X^9
\label{linx9}
\end{eqnarray}
with $u_9=\pm\infty$ solves the equation of motion of the BSFT. This
kink solution connecting the two vacua $T=\pm\infty$ represents a BPS
D8-brane localized in the $x^9$ direction and in fact its tension is
precisely reproduced in the analysis of the super BSFT \cite{KMM2}  
(for the bosonic case and the D-$\overline{\mbox D}$ case, see
\cite{KMM} and \cite{KrLa} respectively).   

Recent development on the tachyon condensation is mostly on its
time-dependent process whose study was initiated in \cite{originalroll}
with use of boundary deformations of conformal field theories, of the
type $T(X)\sim e^{X^0}$, which is called rolling tachyon. An interesting 
outcome was that at late time there remains the ``tachyon matter'' with
finite energy but vanishing pressure. Description of the tachyon matter
in the super BSFT was provided \cite{ST,Minahan} where the tachyon
profile is linear, 
\begin{eqnarray}
 T = u_0 X^0 \ .
\label{lint}
\end{eqnarray}
(Note that this looks quite different from the boundary deformation
above.) Taking 
$u_0 \rightarrow \pm 1$, 
one finds that only the late time limit ($x^0 \rightarrow \infty$) of
(\ref{lint}) is a solution of equations of motion for the BSFT
lagrangian derived for linear tachyon profiles. There is a small
correction to (\ref{lint}) which vanishes exponentially at the late
time. The pressure computed from the super BSFT action with the above
solution vanishes at the late time limit.\footnote{We provide a
description of the tachyon matter in bosonic BSFT in Appendix A.}

At the late time of the rolling tachyon, there should be no degree of
freedom of propagating open strings, since the brane should have been
annihilated by the tachyon condensation. To show this is the aim of this
article.  In \cite{tacmat}, a tachyon  effective field theory whose
solution describes the tachyon matter was introduced, and it was shown
that around that solution there is no plane wave fluctuation. Following
it, \cite{Ue} showed that this is also the case for gauge fluctuations
in the BSFT action of \cite{UeTe} with the above linear tachyon
profiles.  An essential feature was found in \cite{GHY} that in fact the
propagation of the fluctuations in the BSFT action derived with the
linear tachyon profiles is subject to a collapsed light cone given by
{\it Carrollian contraction} of Lorentz group in which the speed of
light becomes zero.  But the problem is that, the BSFT action used in
\cite{GHY} is valid only for the linear profiles, while generic
fluctuations are not of this type. 

In this article, we show the collapse of the light cone of all the
excitations of the open strings. We allow arbitrary fluctuations around
the background (\ref{linx9}) or (\ref{lint}), in the scheme of BSFT. 
The light cone structure can be read off from two-point functions in
BSFT. For superstrings, the BSFT action is just a partition function 
of the worldsheet theory with generic boundary interactions \cite{ZS},  
and in this sense the two-point function for gauge fields in BSFT
was first computed in \cite{Tseytlin-mobius}. More systematically the
tachyon two-point function was computed in the appendix of \cite{KMM},
and also in \cite{Frolov}. (We will explore the two- and three-point
functions in BSFT and its relation to the tachyon condensation, in our
forth-coming paper \cite{HT}.)

For example, the two-point function for the gauge fields in super BSFT
can be computed in the following manner. The worldsheet boundary term
for the gauge field is written as  
\begin{eqnarray}
 I_{\rm B} = -i\int_{\p \Sigma}\!\!
d\tau
 \int\!\! dk \left(
a_\mu(k):\dot{X}^\mu e^{ik_\nu X^\nu}\!:  -2f_{\mu\nu}(k)
:e^{ik_\rho X^\rho}\!:\; \psi^\mu\psi^\nu
\right)
\ ,
\label{gaugev}
\end{eqnarray}
where $a_\mu(k)$ is the momentum representation of the target space
gauge field, $A_\mu(x)=\int\!dk\; a_\mu(k)e^{ikx}$, and
$f_{\mu\nu}(k)=ik_\mu a_\nu - i k_\nu a_\mu$ is that of the field
strength. The two point function of the gauge fields in the target space
is just  $\int\! dx  \langle I_{\rm B} I_{\rm B}\rangle$,  which
can be evaluated with use of worldsheet propagators on the boundary of a
unit disk.  For the flat background with no tachyon condensation, they 
are\footnote{The worldsheet action we use here is 
\begin{eqnarray}
 \frac{1}{4\pi}\int\! d^2z 
\left[
\frac{2}{\alpha'}\p_z X^\mu \p_{\bar{z}} X_\mu 
+ \psi^\mu \p_{\bar{z}} 
\psi_\mu + \tilde{\psi}^\mu \p_z \tilde{\psi}_\mu 
\right] \ .
\end{eqnarray}
The spacetime metric is taken to be $\mbox{diag}(-1,1,1,\cdots,1)$,
and we put $\alpha'=2$ in this article.
} 
\begin{eqnarray}
 \left\langle \hat{X}^\mu(\tau) \hat{X}^\nu(0)\right\rangle
= -4 \eta^{\mu\nu}\log \left|
2\sin\frac{\tau}{2}
\right| \ , \quad 
\left\langle \psi^\mu(\tau) \psi^\nu (0)\right\rangle
 = \frac{1}{2\sin\frac{\tau}{2}} \ ,
\label{norpro}
\end{eqnarray}
where the zero mode for $X$ is already subtracted: 
$X^\mu = x^\mu + \hat{X}^\mu$. A straightforward calculation of
$\int\! dx  \langle I_{\rm B} I_{\rm B}\rangle$ exhibits
the on-shell condition \cite{HT}. Massive excitations can be treated in
the same manner, at least for two-point functions in the BSFT. We apply
this strategy to extract the light cone structure of the open string
excitations in the background linear profiles of the tachyon. Since we
are interested in small fluctuations around the backgrounds, the
two-point functions in the BSFT contain enough information on our
concern. The open string excitations we treat here are of the standard
normalization in no background coupling, and we would like to see how
they behave once they are in the nontrivial tachyon background. What we
shall find is that peculiar behavior of the worldsheet propagators, such
as vanishing / diverging, results in the death of open strings ---
singular structures of the light cones. 


\section{Death of open strings}

\subsection{Spatial tachyon kink and descent relation}

As a warm-up, we consider a spatial kink of a linear profile of the
tachyon in a non-BPS D9-brane (\ref{linx9}). In this background which is
off-shell except $u_9=0$ or $\pm\infty$, the relevant worldsheet
boundary propagator is written as \cite{KMM2} 
\begin{eqnarray}
\left\langle X^9(\tau) X^9(0)\right\rangle 
= 2\sum_{m\in {\bf Z}} \frac{1}{|m|\!+\!u_9^2} e^{im\tau}
\ , \quad 
\langle \psi^9(\tau) \psi^9(0)\rangle
=-\frac{i}{2} \!\!\sum_{r\in {\bf Z}+1/2} 
\frac{r}{|r|\!+\!u_9^2} e^{ir\tau} 
 \ . \hspace{8mm}
\label{x9co}
\end{eqnarray}
Here we included the zero mode in the propagators. When we take the
$u_9=\pm\infty$ solution representing a BPS D8-brane, the correlators
for $X^9$ and $\psi^9$ vanish completely. Let us consider this effect on
the two-point functions in the super BSFT in detail. First, the $k^9$
dependence in the resultant two-point functions in BSFT disappears. This
is because the momentum $k^9$ is always coupled to $\hat{X}^9$ and so
its contractions in $:e^{ik\cdot\hat{X}}\!:$ disappear when 
$|u_9| \rightarrow \infty$. The momentum representation of the BSFT
action for the fluctuations is independent of $k_9$, which means that 
in the coordinate representation the action doesn't contain $\p_9$
and especially any kinetic term along the $x^9$ direction. Therefore, in
this background, any excitation of an open string cannot propagate in
the $x^9$ direction. The relevant light cone collapses to the shape of a
fan: the effective inverse metric appearing in the kinetic term becomes 
\begin{eqnarray}
G^{\mu\nu} = 
 \lim_{u_9 \rightarrow \infty}
\mbox{diag} \left(-1,1,\cdots,1,{\cal O}(1/u_9^2)\right) \ .
\end{eqnarray}
Physics at different values of $x^9$ are decoupled from each other.

An additional fact is that for this background the zero mode integral in
the partition function gives a localization of the worldvolume
\cite{BSFT,KMM,KMM2}, that is, an overall delta function $\delta(x^9)$
in front of the BSFT lagrangian. This means that the propagation only at
the selected value $x_9=0$ survives, while the restricted propagation at
$x_9\neq 0$ has a trivial vanishing action. Therefore, the gauge fields
exist only at the D8-brane worldvolume $x^9=0$ and the propagation is
only along the D8-brane, which is consistent with the Sen's conjecture. 

Furthermore, for the massless gauge bosons for example, the polarization 
parallel to $X^9$ should disappear in this background since the 
resultant brane has no worldvolume direction for that. This in fact
occurs, since the kinetic term for this polarization comes from the 
contraction of $\dot{X}^9$ or $\psi^9$ in the vertex operators of the
gauge fields (\ref{gaugev}). 

The same mechanism is applicable to massive excitations, and all the
excitations are subject to these constraints. They cannot propagate
along $x^9$, they live at $x^9=0$, and the polarizations are restricted
in such a way that the worldvolume theory is just that of the 8+1
dimensions. We have seen here that the vanishing of the worldvolume
propagator results in the collapse of the light cone for the open string
excitations.\footnote{In the renormalization group approach
\cite{HaKuMa} or the boundary state approach \cite{AsSuTe}, it was shown  
that the Neumann boundary condition for the open string turned to the
Dirichlet boundary condition due to the tachyon condensation. This is
also consistent with our result.}

\subsection{Rolling tachyon}

Physics in the background of tachyon profile linear in time is, as we
shall see, different from the above static case. In \cite{ST,Minahan}, a
time-dependent spatially-homogeneous tachyon decay was analysed in the
scheme of BSFT, by adopting the linear tachyon profile (\ref{lint}). 
This profile is on-shell for $|u_0|=1$ only at the late time limit, 
$x^0 \rightarrow \infty$, in the following sense: the BSFT equations of
motion derived with the linear profile has a solution which differs from
(\ref{lint}) by a small correction dumping exponentially in time. 
Writing $u_0^2=1-\kappa$ with a small positive parameter $\kappa$, 
one obtains at the late time $\kappa\propto e^{-(x^0)^2/8}$ \cite{ST}.
(Although this explicit form of $\kappa$ may be corrected, we expect
$\kappa \rightarrow 0$ in the late time behavior of the rolling tachyon 
as disscussed in \cite{ST}.) This is the rolling tachyon in the super
BSFT. For bosonic strings, see Appendix A. The correction $\kappa$ is
time-dependent, but at late time its time-dependence becomes very small
and we may treat this as a constant perturbation from $|u_0|=1$.

The nature of the propagating fluctuations in this late-time rolling
tachyon background is encoded in the BSFT two-point functions. We
evaluate a super BSFT two-point function for the gauge field excitation
of an arbitrary profile (\ref{gaugev}), around this
background.\footnote{The reason why we use gauge excitations first is
that we know $A_\mu=0$ is on-shell around the rolling tachyon
(\ref{lint}). Treatment of the tachyon fluctuation needs a precaution
because the background (\ref{lint}) is on-shell only in the late time
limit.} It is enough to consider the two-point functions since we are
interested in a small fluctuation of a given open string field. The
kinetic structure of the fluctuations may largely depend on the
worldsheet propagators, as we have seen in the previous subsection.  So
let us look closely at the worldsheet propagators. The propagators in
this background include  
\begin{eqnarray}
\langle X^0(\tau) X^0(0)\rangle
=-2 \sum_{m\in {\bf Z}} \frac{1}{|m|-u_0^2} e^{im\tau } \ ,
\;\;
\langle \psi^0(\tau) \psi^0(0)\rangle
=\frac{i}{2} \!
\sum_{r\in {\bf Z}+1/2} \frac{r}{|r|-u_0^2} e^{ir\tau } \ .
\label{corsup}
\;\;\;\;\;\;
\end{eqnarray}
Note that the sign in front of $u^2$ is different from that of the
previous subsection due to the sign of the metric. Consequently, the
behavior of the correlators is quite different from that of
(\ref{x9co}). Let us expand the correlator for $X$'s in (\ref{corsup})
for a small $\kappa$. In (\ref{corsup}), first the term with $m=\pm1$  
is the leading order and in fact divergent in the limit 
$\kappa \rightarrow 0$. The next-to-leading order term is given by 
effectively shifting $m$ by one in the summation, thus we obtain the
expansion 
\begin{eqnarray}
 \left\langle \hat{X}^0(\tau) \hat{X}^0(0)\right\rangle
= -2\left[
\frac{1}{\kappa} (e^{i\tau}\! +\! e^{-i\tau})
-e^{i\tau}\log(1\!-\!e^{i\tau})
-e^{-i\tau}\log(1\!-\!e^{-i\tau})
+ {\cal O}(\kappa)
\right]  . \;\;\;
\label{propx0}
\end{eqnarray}
Note here that we have already subtracted the zero mode part, $m=0$.
As we shall see, this divergence is related to the collapse of the
light cone. 

For this $X^0$ direction, the change in the propagator forces us to
redefine the normal ordering of operators. Denote the normal ordering
with the propagator (\ref{corsup}) as $ \no e^{ik_\mu \hat{X}^\mu}\no $,  
then the relation to the usual normal ordering with the propagator
(\ref{norpro}) is   
\begin{eqnarray}
 :e^{ik_\mu \hat{X}^\mu}: &=& \exp\left[
\frac12 (k_0)^2 \left(
\left\langle \hat{X}^0(\epsilon)\hat{X}^0(0)
\right\rangle_{u_0=0, \epsilon=0}-
\left\langle \hat{X}^0(\epsilon)\hat{X}^0(0)
\right\rangle_{u_0, \epsilon=0}
\right)
\right]
 \no e^{ik_\mu \hat{X}^\mu}\no   
\nonumber
\\
& = & \exp\left[
(k_0)^2\sum_{0\neq m\in {\bf Z}} \frac{u_0^2}{|m|(|m|-u_0^2)}
\right]
 \no e^{ik_\mu \hat{X}^\mu}\no   \ . 
\end{eqnarray}
This momentum-involved redefinition can be evaluated for the small
$\kappa$ as  
\begin{eqnarray}
 :e^{ik_\mu \hat{X}^\mu}\!: &=& 
\exp\left[
\frac{2}{\kappa} (k_0)^2  - \frac{\pi^2\kappa}{3}(k_0)^2 
+ {\cal O}(\kappa^2)
\right]
 \no e^{ik_\mu \hat{X}^\mu}\no   \ .
\label{renorm}
\end{eqnarray}
This additional factor can be absorbed into the definition of
$a_\mu(k)$, if one wants. But we leave it because we would like to see 
what happens to the gauge field with the standard normalization defined
with the normal ordering $:e^{ik_\mu \hat{X}^\mu}\!:$.

The gauge $A_0=0$ reduces the computation of the two-point function,
\begin{eqnarray}
  \langle I_{\rm B} I_{\rm B}\rangle
= -\int\!\! dk d\tilde{k} 
\int \frac{d\tau}{2\pi}
\left[
a_i(k) a_j (\tilde{k})
\left(\frac{-\delta_{ij}}{\sin^2\frac{\tau}{2}}
-4\tilde{k}_i k_j \cot^2\frac{\tau}{2}
\right)
+f_{ij}(k)f_{ij}(\tilde{k})
\frac{-1}{\sin^2\frac{\tau}{2}}
\right.
\nonumber \\
\left.
-8 f_{i0}(k)f_{i0}(\tilde{k})\frac{1}{\sin\frac{\tau}{2}}
\langle \psi^0(\tau)\psi^0(0)\rangle
\right]
\langle:e^{ikX}(\tau): :e^{i\tilde{k}X}(0):\rangle \ .
\nn
\end{eqnarray}
The correlator for the directions other than $X^0$ can be evaluated
easily with (\ref{norpro}) as 
\begin{eqnarray}
 \left\langle :e^{ik_i X^i(\epsilon)}: :e^{i\tilde{k}_i X^i(0)}: 
\right\rangle 
= |1-e^{i\epsilon}|^{4k_i\tilde{k}_j\delta^{ij}}
e^{i(k_i+\tilde{k}_i)x^i} \ . 
\end{eqnarray}
Using the expansion (\ref{propx0}), we can include the direction $X^0$
and obtain 
\begin{eqnarray}
\lefteqn{
\iint \frac{d\tau_1 d\tau_2}{(2\pi)^2}
\left\langle \no e^{ik_\mu \hat{X}^\mu(\tau_1)}\no  
\no e^{i\tilde{k}_\mu \hat{X}^\mu(\tau_2)}\no  
\right\rangle
}
\nonumber \\
&& =
\int_0^{2\pi} \frac{d\tau}{2\pi} |1-e^{i\tau}|^{4k_i \tilde{k}_i} 
e^{\frac4{\kappa} k_0 \tilde{k}_0 \cos\tau}
(1-e^{i\tau})^{-2k_0\tilde{k}_0e^{i\tau}}
(1-e^{-i\tau})^{-2k_0\tilde{k}_0e^{-i\tau}}e^{{\cal O}(\kappa)}
\ . 
\;\;\;\;\;
\end{eqnarray}
To evaluate this, we assume $k_0 \tilde{k}_0<0$.\footnote{Now the
background is time-dependent, so the momentum is not conserved, 
$k_0 + \tilde{k}_0\neq 0$ generically. The case of 
$k_0 \tilde{k}_0 > 0$ can be treated in a similar manner but with
a steepest-descent method. }
The integrand has a sharp peak at $\cos \tau \sim -1$ when $\kappa$ is
very small. We may treat it as a delta function with an appropriate
normalization dependent on $\kappa$. For a finite $k_0 \tilde{k}_0$ 
and in the limit $\kappa \rightarrow 0$, we obtain a formula for any
$\kappa$-independent smooth function $g(\tau)$,  
\begin{eqnarray}
 \int\!\! \frac{d \tau}{2\pi}\; g(\tau)
\langle \; \no e^{ik\hat{X}}(\tau)\no 
\no e^{i\tilde{k}\hat{X}}(0)\no \rangle
= g(\pi)
\sqrt{\frac{-\kappa}{8\pi k_0 \tilde{k}_0}}
\exp\left[
-\frac{4}{\kappa}
k_0 \tilde{k}_0 + 4\log 2 k_i \tilde{k}_i
\right] \ . \;\;\;\;
\end{eqnarray}
Then finally the nonzero-mode part of the two-point function reads
\begin{eqnarray}
   \langle I_{\rm B} I_{\rm B}\rangle
= \int\!\! dk d\tilde{k} 
\left[
a_i (k)a^i(\tilde{k})+2 f_{ij}(k)f^{ij}(\tilde{k})
-4 f_{i0}(k) f^{i0}(\tilde{k})\left(3+\pi\right)
\right]
\nonumber \\
\sqrt{\frac{-\kappa}{8\pi k_0 \tilde{k}_0}}
\exp\left[
-\frac{4}{\kappa}
k_0 \tilde{k}_0 + 4\log 2 k_i \tilde{k}_i
+ \frac{2}{\kappa}(k_0^2 + \tilde{k}_0^2)
\right] \ .
\;\;\;\;\;
\end{eqnarray}
In this expression, the limit $\kappa \rightarrow 0$ gives a divergent
factor, therefore this shows that any nonzero $k_0$ results in no
dynamics. For any non-vanishing $k_0$, the two point function diverges 
and then if this serves as a kinetic term in a quantum field theory the
path integral with non-vanishing $k_0$ is highly oscillatory, with which
any correlation function vanishes.\footnote{Or saying it differently,
the magnitude of the BSFT kinetic term roughly corresponds to an
effective ``tension'' of the brane felt by the string excitations. In
the present case this diverges and resultantly the open string modes
effectively disappear since it is too heavy to excite.}
\setcounter{footnote}{0}

For the massive excitations, the same physics applies, which implies
that there is no propagation of open strings in the rolling tachyon
background at the late time. This is simply because all the open string
excitations are accompanied with the momentum eigenfunction
$:e^{ik\cdot\hat{X}}:$ whose contraction necessarily gives the
divergence whenever it has a non-vanishing $k_0$. 

Thus, whole the open string excitations are subject to a light cone of 
the so-called Carrollian contraction of Lorentz group
\cite{GHY}.\footnote{ 
Let us briefly mention the relation to the computations in \cite{GHY}
where the BSFT action \cite{UeTe} derived by the linear tachyon
background was adopted. In our respect, the computations in \cite{GHY}
corresponds to a different limit, that is, taking $k_0 \rightarrow 0$
first and then $\kappa \rightarrow 0$, since there only the constant
gauge field strength was considered.} The light cone becomes a singular
half-line and no open string can move on the original worldvolume of the
brane. The upper limit of the speed of the propagations becomes zero.
This describes how the open strings die in the late-time rolling
tachyon. 

It was suggested from the effective field theory that the spatial
distribution of the open string condensations becomes arbitrary, and so
there appears a huge degeneracy in the spatial configurations
\cite{GHY}.  This can also be seen in the above result. Since the 
dependence on the spatial momentum $k_i \tilde{k}_i$ is in the 
next-to-leading order in $\kappa$, we can always set the small 
${\cal O}(\kappa)$ dependence in $k_0$ in such a way that it cancels
arbitrary function of the sub-leading order written by $k_i$. Therefore
in the limit $\kappa \rightarrow 0$, any dependence of $k_i$ can be
absorbed into the vanishing $k_0$ and so allowed. This is the reflection
of the fact that there is no interaction between different points on the
original brane worldvolume, in the late-time limit of the rolling
tachyon. 

\subsection{Rolling tachyon in background string charge}

We may extend this relation between the divergence in the worldsheet
propagators and the light cone structure of the excitations on the
decaying brane, to the situation with a background string charge on the
D-brane \cite{bE}. This is the setup used for the tree level open-closed
duality stated in \cite{tree}. The duality statement is supported by a
Nambu-Goto analysis \cite{Yi} in which the string oscillations
can propagate with a reduced speed of light proportional to the
background string density $E$. This reduction of the speed was first
observed in a low energy effective field theory \cite{GHY}. Indeed, we
will see in the BSFT two point function that the light cone structure in
the rolling tachyon with the presence of this electric field $E$ is just
that. So, in this case, the open string excitations don't cease to
propagate but a propagation along the background electric field
$\vec{E}$ is allowed with the reduced speed.  

In the background electric field $F_{01}\equiv E$ with the tachyon
profile linear in time, the worldsheet propagator is given by
\cite{li-witten} 
\begin{eqnarray}
 \left\langle \hat{X}^\mu(\tau) \hat{X}^\nu(0)\right\rangle
=\sum_{m=1}^\infty \left(
M_+^{\mu\nu} e^{im\tau}
+M_-^{\mu\nu} e^{-im\tau}
\right) \ , 
\end{eqnarray}
where the nontrivial parts of the matrices $M_\pm$ are
\begin{eqnarray}
&& M_+ 
= \left(
\begin{array}{cc}
-m\!+\!u^2_0 & E\\ -E & m
\end{array}
\right)^{-1}
= \frac{1}{m(m-u^2_0)-m^2 E^2}
\left(
\begin{array}{cc}
-m  & mE\\ -mE & m\!-\!u_0^2
\end{array}
\right) \ ,
\\
&& M_- 
= \left(
\begin{array}{cc}
-m\!+\!u^2_0 & -E\\ E & m
\end{array}
\right)^{-1}
= \frac{1}{m(m-u^2_0)-m^2 E^2}
\left(
\begin{array}{cc}
-m  & -mE\\ mE & m\!-\!u^2_0
\end{array}
\right) \ . 
\end{eqnarray}
Our concern is the divergence appearing for $m=1$, at 
$u_0 = \pm \sqrt{1-E^2}$ which is the rolling speed of the tachyon
\cite{bE,GHY}. We perturb it as before, $u_0^2 = 1-E^2 -\kappa$ with a
small positive $\kappa$, then the divergent parts of the relevant
propagators are 
\begin{eqnarray}
&& \langle X^0(\tau) X^0(0) \rangle
= \frac{-4}{\kappa}\cos \tau
\ , \quad 
\langle X^1(\tau) X^1(0) \rangle
= \frac{4E^2}{\kappa}\cos \tau \ ,
\\
&& \langle X^0(\tau) X^1(0) \rangle
=-\langle X^1(\tau) X^0(0) \rangle
= \frac{2iE}{\kappa}\sin \tau \ .
\end{eqnarray}
The change of the normal ordering is\footnote{On-shell gauge field in no
tachyon condensation satisfies $k_0^2-k_i^2=0$, so the exponent seen
here is usually positive.} 
\begin{eqnarray}
 :e^{ik_\mu \hat{X}^\mu}: &=& 
\exp\left[
\frac{2}{\kappa} \left(k_0^2  -E^2 k_i^2\right)
+ {\cal O}(1)
\right]
 \no e^{ik_\mu \hat{X}^\mu}\no   \ .
\label{renoe}
\end{eqnarray}
The exponential factor appearing in the partition function from the
correlators among vertex operators reads
\begin{eqnarray}
 \exp\left[
\frac{4}{\kappa}\left(
k_0 \tilde{k}_0 - E^2 k_1 \tilde{k}_1
\right)\cos\tau
- \frac{2iE}{\kappa} \sin\tau
\right] \ . 
\end{eqnarray}
This factor has again a sharp peak at $\tau=\pi$ (when 
$k_0 \tilde{k}_0 - E^2 k_i \tilde{k}_i<0$), thus after an integration
over $\tau$, it results in a factor  
\begin{eqnarray}
 \exp\left[
-\frac{4}{\kappa}\left(
k_0 \tilde{k}_0 - E^2 k_1 \tilde{k}_1
\right)
+\frac{2}{\kappa} \left(k_0^2  -E^2 k_1^2\right)
+\frac{2}{\kappa} \left(\tilde{k}_0^2  -E^2 \tilde{k}_1^2\right)
+{\cal O}(1)
\right] \ , 
\label{expa}
\end{eqnarray}
where we have included the factor coming from (\ref{renoe}). This 
diverges for nonzero $k_0^2 - E^2 k_1^2 $ (or one or all $k$ replaced by
$\tilde{k}$). Thus we obtain a constraint for propagating degrees of
freedom 
\begin{eqnarray}
 k_0^2 - E^2 k_1^2 = {\cal O}(\kappa) \rightarrow 0
\ .
\end{eqnarray}
which shows that only the propagation along $\vec{E}$ can be allowed and
its speed is in fact $E$. This is consistent with the analysis of 
highly-oscillated Nambu-Goto strings \cite{Yi}. 

In \cite{GHY}, this reduction of the speed of light was obtained for low
energy limit of gauge field excitation in the rolling tachyon background.
The computation of \cite{GHY} can be reproduced from our perspective if 
we consider a constant field strength, which was the starting point of
\cite{GHY}, as the low energy limit. This means that one took the limit
$k\rightarrow 0$ before taking $\kappa \rightarrow 0$. Our present
calculation shows that the reduction of the light cone structure appears
not only for the very low energy but also for all the open string
excitations with any momentum. The light cone becomes the shape of a fan,
which can also be read from the effective inverse metric $G^{\mu\nu}$
defined when the exponent of (\ref{expa}) is rewritten as 
\begin{eqnarray}
 -\frac{2}{\kappa} \left(k_\mu - \tilde{k}_\mu\right)
G^{\mu\nu} \left(k_\nu - \tilde{k}_\nu\right)\ ,\quad 
G^{\mu\nu} = \mbox{diag} (-1,E^2, 
\underbrace{0,0,\cdots,0}_{{\cal O}(\kappa)}) \ .
\end{eqnarray}


\section{Summary and discussions}

The essential point of this article is that the divergence or 
disappearance of the worldsheet boundary propagators crucially affects
propagation of string excitations in the target space. The connection
between the worldsheet and the target space is provided in the scheme of
the BSFT. For the linear profile $T = u_9 X^9$ 
($u_9\rightarrow\pm\infty$), the $X^9$ propagator vanishes and
resultantly the 99 component of the target space effective inverse
metric is eliminated. For the rolling tachyon 
$T=u_0 X^0$ ($u_0 \rightarrow \pm 1$), the $X^0$ propagator diverges,
and the spatial component of the target space effective inverse metric 
vanishes. Therefore the information of the target space metric is
encoded in the ratios among various worldsheet propagators.

This feature is in fact found also in the worldvolume of a BPS
D$p$-brane without the tachyon condensation but with a constant gauge
field strength. The effective open string metric of this case
appears as a coefficient of the $\log$ part of the 
boundary propagator of the worldsheet bosons \cite{prop}:
$G^{\mu\nu}=[1/(\eta + F)]^{(\mu\nu)}$ where the indices are
symmetrized. 
First, taking a limit $F_{12}\rightarrow \infty$, we have vanishing 
propagators for $X^1$ and $X^2$, so following the argument in Section
2.1, we observe that the light cone structure of the fluctuations
is reduced to that of a D$(p\!-\!2)$-brane. This is in fact expected,
since in this limit the bound charge of the D$(p\!-\!2)$-branes,
mesured by the magnitude of $F_{12}$, diverges and the system is
expected to be saturated by the collection of the 
D$(p\!-\!2)$-branes. Second, let us consider another limit
$F_{01}\rightarrow 1$. The propagators for $X^0$ and $X^1$ diverge.
We find a light cone similar to that of Section 2.3 but with $E=1$. In
this case the bound fundamental strings whose charge diverges saturate
the brane system so that only propagations along the strings are
allowed. (See \cite{Herdeiro} for a related study of the light cones.)

The Carrollian contraction of the Lorentz group found here is based
on the divergence of the two point functions of standard open string
vertex operators. One can argue that this divergence might be cured by
``renormalization'', that is, a momentum-dependent field redefinition of
the field. In fact, as indicated in (\ref{renorm}), a natural
normalization in this late-time rolling tachyon background may be with
the normal ordering $\no e^{ik\cdot X}\no$. This causes the field
redefinition of the form $a'(k)\equiv\exp[2k_0^2/\kappa] a(k)$ 
($\kappa \rightarrow 0$), but even with this normalization, the
two-point function diverges, due to the contraction among the
correlators: $\exp [-4k_0 \tilde{k}_0/\kappa]$.  

Supposing that $k_0$ and $\tilde{k}_0$ are of the same order, a stronger
renormalization $a'(k)\equiv\exp[4k_0^2/\kappa] a(k)$ looks to make the
divergent factor eliminated and might give a canonically normalized
kinetic term for the fluctuations. However, more severe divergence can
be found in general $n$-point functions with $n>2$. Since the $n$-point
function includes arbitrary contraction among $n$ operators of the form
$:\!e^{ik_0X^0}\!\!:$, it may diverge much strongly than what can 
be eliminated with the above field redefinition. Thus the
``renormalized'' two-point function becomes rather meaningless. The
resultant theory appears to be a strongly interacting theory and no
perturbative state defined with the renormalized $:e^{ik\cdot X}:$ may
appear as an asymptotic state, unless $k_0=0$. This is consistent with
the picture of confinement of open strings at the tachyon vacuum
\cite{Yi-con}.   

In cubic string field theory \cite{Wittencubic}, various attempts have
been made to understand the open string excitations at the true vacuum
\cite{HataTera,TZ,drukker}. These attempts are for static homogeneous
tachyon condensation in which the constant tachyon is put to its true
vacuum. Thus a direct relation to our BSFT approach is unclear. For the
constant tachyon in BSFT, the worldsheet boundary receives a weight
factor $\sim e^{-T^2}$ which results in the vanishing of the worldvolume
D-brane action, while in CSFT similar structure was found \cite{TZ}. Due
to this boundary factor, the boundary shrinks effectively and disappear
\cite{KMM}, which seems to be realized also in CSFT \cite{drukker}. 
Further concrete relations among the two SFT's should be clarified in
the future.  

The rolling tachyon process has been considered much in deformed
conformal field theories, especially boundary Liouville theories 
\cite{stro,bl}. It was pointed out in \cite{fred} that the late-time
spectrum has vanishing wave functions while its $k_0$ dependence is
diverging, which looks quite consistent with our results of the
divergence of the two-point functions. It is interesting to seek for a
relation of our BSFT approach to the creation of open strings due to the
change of the vacua in the conformal field theories \cite{stro}.


\acknowledgments 

K.H.\ would like to thank M.~Kato and P.~Yi for valuable discussions. 
S.T.\ would like to thank A.~Parnachev, S.~Sugimoto and T. Takayanagi
for useful discussions. 
The work by K.H. 
was supported in part by the
Grant-in-Aid for Scientific Research (No.~12440060, 13135205, 15540256
and 15740143) from the Japan Ministry of Education, Science and Culture.

\appendix

\section{Tachyon matter in bosonic BSFT}

In this appendix, we present a description of tachyon matter in bosonic
string theory, as a solution of a bosonic BSFT. This is an analogue of
\cite{ST,Minahan} given for superstring.  
The analysis presented in this article for the light cone structure is
applicable to the bosonic strings in the same manner.

\vspace{5mm}
\noindent
\underline{BSFT action}

First, let us construct a BSFT action by closely following the original
construction \cite{BSFT} but slightly generalizing it to include a
time-like target-space direction $X^0$. We introduce a tachyon profile at
the boundary of the worldsheet,  
\begin{eqnarray}
 S_{\rm bdry} = \int\! \frac{d\theta}{2\pi}\; T(X)
\ , \quad
T(X)\equiv 
a+\sum_{i=1}^{25} \frac{u_i}{4} (X^i)^2 - \frac{u_0}{4}(X^0)^2
 \ ,
\label{sbdry}
\end{eqnarray}
where $a, u_i$ and $u_0$ are boundary deformation parameters. We have
defined this with the minus sign in front of the $X^0$ mass term so that
the propagators for $X$'s have the following common form: 
\begin{eqnarray}
 \left\langle (X^i)^2(\theta) \right\rangle
= \frac{2}{u_i}-\sum_{m=1}^\infty   \frac{4u_i}{m(m\!+\!u_i)} \ ,
\quad
 \left\langle (X^0)^2(\theta) \right\rangle
= -\left(
\frac{2}{u_0}-\sum_{m=1}^\infty   \frac{4u_0}{m(m\!+\!u_0)}
\right) \ .
\nn
\end{eqnarray}
Following the definition of the partition function given with 
(\ref{sbdry}),  we find 
\begin{eqnarray}
\frac{d}{d u_i} \log Z_1 = -\frac{1}{8\pi} \int_0^{2\pi}d\theta
\left\langle
(X^i)^2(\theta)
\right\rangle \ , 
\quad \frac{d}{d u_0} \log Z_1 = \frac{1}{8\pi} \int_0^{2\pi}d\theta
\left\langle
(X^0)^2(\theta)
\right\rangle \ .
\nn
\end{eqnarray}
Obviously the partition function is a product of a function of $u_i$ and
that of $u_0$, so we obtain (up to a constant overall factor)
\begin{eqnarray}
 Z = e^{-a}Z_1(u_0)\prod_{i=1}^{25}Z_1(u_i) \ , 
\quad 
 Z_1(u) \equiv \sqrt{u} e^{\gamma u} \Gamma(u) \ .
\end{eqnarray}
In the derivation of the action from the partition function, which was
described in \cite{BSFT} in detail, it is easy to find that the
inclusion of $X^0$ is just done by regarding this $X^0$ as $i X^{26}$,
and accordingly $u_0 = u_{26}$. Then
\begin{eqnarray}
S = \left(
1 + a + 
u_0 -u_0 \frac{\partial}{\p u_0}+ 
\sum_{i=1}^{25}
\left(
u_i -u_i \frac{\partial}{\p u_i}
\right)
\right) Z(a,u_0,u_i) \ .
\end{eqnarray}
For our purpose, we turn on only $u_0$ and $u_1$. (Although a
homogeneous rolling tachyon solution is obtained with $u_1=0$, we need
this $u_1$ dependence to evaluate its pressure later.) We need a
lagrangian written in terms of the target space tachyon field $T(x)$
defined in (\ref{sbdry}). Using the following formulas 
\begin{eqnarray}
 \int\! dx^1 dx^0 \; e^{-T} = e^{-a}\frac{4\pi}{\sqrt{-u_0u_1}}
\ , \quad 
 \int\! dx^1 dx^0 \; T e^{-T} = e^{-a}\frac{4\pi}{\sqrt{-u_0u_1}}(a+1)
\ , 
\end{eqnarray}
we obtain the lagrangian as 
\begin{eqnarray}
 L = e^{-T}
\frac{\sqrt{-u_0u_1}}{4\pi}
\left[
T + 
u_0 -u_0 \frac{\partial}{\p u_0}+ 
u_1 -u_1 \frac{\partial}{\p u_1}
\right] Z_1(u_0)Z_1(u_1) \ .
\label{lag}
\end{eqnarray}
We may replace the parameters $u$ by 
\begin{eqnarray}
 u_1 \rightarrow 2\p_1 \p_1 T, \quad 
 u_0 \rightarrow -2\p_0 \p_0 T
\end{eqnarray}
which follow from the definition of the tachyon profile (\ref{sbdry}).
With the simple substitution of these expression to (\ref{lag}), we
obtain a lagrangian written solely by the tachyon field. Note that in
this lagrangian the covariance is not manifest, since we have suppressed
the dependence on $\p_1 \p_0 T$ in the computation. However, the Lorentz
invariance can be restored in the following way. The lagrangian
(\ref{lag}) can be Taylor-expanded in terms of $u_0$ and  $u_1$. The
first nontrivial term comes like $ u_0 + u_1$ times a factor of a
function of $T$ such as $e^{-T}$. We can rebuild its Lorentz-invariant
form 
\begin{eqnarray}
 u_0 + u_1 = 2 \eta^{\mu\nu}\p_\mu \p_\nu T \ . 
\end{eqnarray}
So we might expect that the lagrangian can be written only by this
invariant, as is usually used for BSFT lagrangians for
superstrings.\footnote{ 
Note that even for superstrings we face a similar problem 
when we consider D-$\overline{\mbox{D}}$ or several D-branes \cite{Te}.}  
However this is not the case. The next-to-leading term has the following
form 
\begin{eqnarray}
 f(T)\left(
(u_0)^2 + (u_1)^2
\right)+ g(T)u_0 u_1 \ .
\end{eqnarray}
This is lifted to its Lorentz-invariant expression,
\begin{eqnarray}
4 f(T) 
\eta^{\mu\nu}\eta^{\rho\sigma}\p_\mu \p_\rho T \p_\nu \p_\sigma T
 - 2 g(T)\left( (\eta^{\mu\nu}\p_\mu\p_\nu T)^2 - 
\eta^{\mu\nu}\eta^{\rho\sigma}\p_\mu \p_\rho T \p_\nu \p_\sigma T
\right) \ .
\end{eqnarray}
Higher order terms include much more intricate Lorenz contractions but
they can be uniquely lifted to their Lorentz-invariant expressions.  

\vspace{5mm}
\noindent
\underline{Rolling solution}

Let us find a time-dependent homogeneous solution. Instead of solving an
equation of motion, homogeneous time-dependent solution can be easily
obtained by solving the energy-conservation condition. The easiest way
to get the hamiltonian from the above lagrangian (\ref{lag}) is to
couple the system to a background metric and derive the
energy-momentum. In the present case all the nontrivial Lorentz 
contractions reduce to the simple single invariant 
$u_0 = -2(v\p_0 (v\p_0 T))$ where $v \equiv \sqrt{-g^{00}}$. Our action
is written as  
\begin{eqnarray}
 S = \int\! dx^0\; \frac{1}{v} e^{-T} (F(u_0) + T G(u_0))
\end{eqnarray}
where
\begin{eqnarray}
F(u_0) \equiv \sqrt{-u_0} \left(
\frac12 + u_0 - u_0 \frac{\p}{\p u_0}
\right)Z_1(u_0)
\ , \quad 
G(u_0) \equiv \sqrt{-u_0}Z_1(u_0) \ . 
\end{eqnarray}
We find the hamiltonian 
$H = -[{\p S}/{\p v}]_{v=1}$ as 
\begin{eqnarray}
H = e^{-T} (F\! +\! TG)- 2 \p_0 \left(
\p_0 T e^{-T}(F'\! +\! T G')
\right)+4\p_0^2 T 
\left(
e^{-T}(F'\! +\! T G')
\right) \ . 
\label{ham}
\end{eqnarray}

The tachyon matter is defined as the late time residual matter when the
tachyon rolls to its true vacuum $T=\infty$. So we are interested in the 
late time behavior of the system given by the above hamiltonian
(\ref{ham}). First, $u_0$ should be negative so that the tachyon may not
roll down in the wrong side of the potential. The hamiltonian should be
conserved, while the overall exponential factor in the above expression
is vanishing as $e^{-T} \sim \exp[u_0 (x^0)^2/4]$. So we expect that the
other factors in the hamiltonian diverge. In fact, from the explicit
expression for $Z_1$, we find, at $u_0 \sim -1$, 
\begin{eqnarray}
 F(u_0) \sim  (1 + u_0)^{-2} \ , \quad 
 G(u_0) \sim  (1 + u_0)^{-1} \ .
\end{eqnarray}
The most diverging factor comes from the second term in (\ref{ham}), 
\begin{eqnarray}
 H \sim -2 \p_0 T e^{-T}(-2\p_0^3T) F'' \ .
\end{eqnarray}
We consider a small fluctuation around the tachyon profile of $u_0=-1$, 
\begin{eqnarray}
 T = a + \frac{1}{4}(x^0)^2 + \epsilon(x^0) \ .
\end{eqnarray}
Then the energy conservation implies
\begin{eqnarray}
x^0 e^{-(x^0)^2/4} (\p_0^3 \epsilon ) (\p_0^2\epsilon)^{-4} 
= \mbox{const.}
\end{eqnarray}
This can be solved as 
\begin{eqnarray}
\ddot{\epsilon} \sim (x^0)^{2/3} \exp [-(x^0)^2/12].
\label{sol}
 \end{eqnarray}
So the fluctuation vanishes in the late time limit, which shows the
consistency of our ansatz.

\vspace{5mm}
\noindent
\underline{Pressure}

Finally we would like to show that the pressure of the homogeneous
solution obtained here is exponentially dumping. The pressure along
$x^1$ direction should be defined as 
\begin{eqnarray}
 p_1 = 2 \frac{\p L}{ \p g^{11}}.
\end{eqnarray}
To this end we have to work with nonzero $u_1$ since in it nontrivial
dependence on $g^{11}$ might appear in the covariant derivatives. As
seen above, the whole structure of the 
Lorenz contraction is very nontrivial. But here we
know that in the end we put the metric to be constant and also the
tachyon to be dependent only on time, which results in a lot of
simplification. The covariant derivatives appearing in the expression is 
\begin{eqnarray}
 \nabla_\mu \p_\nu T = \p_\mu \p_\nu T - \Gamma^{\rho}_{\nu\mu} 
\p_\rho T 
\end{eqnarray}
where the Christoffel symbol is defined as usual, 
$ \Gamma^\rho_{\nu\mu} = (1/2) g^{\rho\sigma}(\p_\mu 
g_{\sigma\nu}\!+\!\p_\nu g_{\sigma\mu}\! -\! \p_\sigma g_{\mu\nu})$.
The index $\rho$ will be contracted with $\p_\rho T$ and so it becomes
irrelevant unless $\rho=0$. We want to see the dependence on $g^{11}$,
then the only term which we find relevant is  
\begin{eqnarray}
 \Gamma^{0}_{11} = \frac{1}{2} \p_0 (1/g^{11}).
\end{eqnarray}
As well as the Christoffel symbols, the metrics used for contracting the
indices of the derivatives can be relevant in the calculation of the
energy-momentum tensor, but this contribution turns out to be
vanishing. This is because after the differentiation with respect to
$g^{11}$ the remaining term always include $\p_1$ which is zero when 
acting on $T(x^0)$. Thus the covariant derivative $\nabla_1$ 
can appear only once in the computation of the pressure, so we need only
a linear term in $u_1$. 

In the evaluation of the pressure, we may replace $u_1$ by 
\begin{eqnarray}
 u_1 \sim -\p_0(1/g^{11}) \p_0 T.
\end{eqnarray}

The lagrangian can be written as, after a little calculation, 
\begin{eqnarray}
 L = \frac{1}{4\pi} e^{-T} \sqrt{-u_0} 
\left(
\frac12 + T + u_0 - u_0 \frac{\p}{\p u_0} + u_1
\right)Z_1(u_0) + {\cal O}(u_1^2) \ .
\end{eqnarray}
so, restoring the $g^{11}$ dependence including the overall factor 
$\sqrt{-g}\sim (g^{11})^{-1/2}$, we obtain
\begin{eqnarray}
 p_1 = - L - 2\p_0 \left(
\p_0 T L
\right) \ .
\label{pres}
\end{eqnarray}
The first term comes from $\sqrt{-g}$ while the second term comes from
the Christoffel in $u_1$. Writing the lagrangian in terms of $F$ and $G$
again, we find that the most dominant part in the pressure comes from 
\begin{eqnarray}
 -\frac{1}{2\pi}\p_0T e^{-T} \p_0 F \; = \;
 \frac{1}{\pi}\p_0T e^{-T} \p_0^3 T  F'  
\; \sim \; \exp[-(x^0)^2/6] \ .
\end{eqnarray}
In the last part we substituted the solution (\ref{sol}). This shows
that the pressure is vanishing at the late time of the rolling tachyon. 
Technically speaking, the reason why the hamiltonian is conserved while
the pressure is dumping is that, in the expressions in terms of $F$, the
number of the derivatives acting on $F$ is less by one in the pressure
compared to that of the hamiltonian. 


\newcommand{\J}[4]{{\sl #1} {\bf #2} (#3) #4}
\newcommand{\andJ}[3]{{\bf #1} (#2) #3}
\newcommand{\AP}{Ann.\ Phys.\ (N.Y.)}
\newcommand{\MPL}{Mod.\ Phys.\ Lett.}
\newcommand{\NP}{Nucl.\ Phys.}
\newcommand{\PL}{Phys.\ Lett.}
\newcommand{\PR}{ Phys.\ Rev.}
\newcommand{\PRL}{Phys.\ Rev.\ Lett.}
\newcommand{\PTP}{Prog.\ Theor.\ Phys.}
\newcommand{\hep}[1]{{\tt hep-th/{#1}}}


\begin{thebibliography}{10}
\bibitem{Senconje}
A.~Sen, {\sl ``Tachyon Condensation on the Brane Anti-Brane System,''}
\J{JHEP}{9808}{1998}{012}, {\tt hep-th/9805170};
{\sl ``Descent Relations among Bosonic D-branes,''}
\J{Int.J.Mod.Phys.}{A14}{1999}{4061}, {\tt hep-th/9902105};
{\sl ``Non-BPS States and Branes in String Theory,''}
{\tt hep-th/9904207};
{\sl ``Universality of the Tachyon Potential,''}
\J{JHEP}{9912}{1999}{027}, {\tt  hep-th/9911116}.

\bibitem{BSFT}
E.~Witten, 
{\sl ``On Background-Independent Open-String Field Theory,''}
\J{Phys.Rev.}{D46}{1992}{5467}, {\tt hep-th/9208027}; 
{\sl ``Some Computations in Background Independent Off-Shell String
	Theory,''} 
\J{Phys.Rev.}{D47}{1993}{3405},  {\tt hep-th/9210065}; \\
S.~L.~Shatashvili, 
{\sl ``Comment on the Background Independent Open String Theory,''}
\J{Phys.Lett.}{B311}{1993}{83}, {\tt hep-th/9303143};
{\sl ``On the Problems with Background Independence in String Theory,''}
{\tt hep-th/9311177}.

\bibitem{Wittencubic}
E.~Witten, 
{\sl ``Noncommutative Geometry and String Field Theory,''}
\J{Nucl.Phys.}{B268}{1986}{253}.\\
See also: 
H.~Hata, K.~Itoh, T.~Kugo, H.~Kunitomo and K.~Ogawa, 
{\sl ``Manifestly Covariant Field Theory of Interacting String,''}
\J{Phys.Lett.}{B172}{1986}{186};
{\sl ``Covariant String Field Theory,''}
\J{Phys.Rev.}{D34}{1986}{2360}.

\bibitem{Sen-nonBPS}
A.~Sen, 
{\sl ``Supersymmetric World-volume Action for Non-BPS D-branes,''}
\J{JHEP}{9910}{1999}{008}, {\tt hep-th/9909062}.

\bibitem{KMM}
D.~Kutasov, M.~Marino and G.~Moore, 
{\sl ``Some Exact Results on Tachyon Condensation in String Field
	Theory,''} 
\J{JHEP}{0010}{2000}{045}, {\tt hep-th/0009148}.

\bibitem{others}
A.~A.~Gerasimov and S.~L.~Shatashvili, 
{\sl ``On Exact Tachyon Potential in Open String Field Theory,''}
\J{JHEP}{0010}{2000}{034}, {\tt hep-th/0009103};
{\sl ''Stringy Higgs Mechanism and the Fate of Open Strings,''}
\J{JHEP}{0101}{2001}{019}, {\tt hep-th/0011009};\\
J.~A.~Minahan and B.~Zwiebach, 
{\sl ``Field theory models for tachyon and gauge field string
	dynamics,''} 
\J{JHEP}{0009}{2000}{029}, {\tt hep-th/0008231};
{\sl ``Effective Tachyon Dynamics in Superstring Theory,''}
\J{JHEP}{0103}{2001}{038}, {\tt hep-th/0009246};
{\sl ``Gauge Fields and Fermions in Tachyon Effective Field Theories,''} 
\J{JHEP}{0102}{2001}{034}, {\tt hep-th/0011226};\\
K.~Hashimoto and S.~Hirano, 
{\sl ``Branes Ending On Branes In A Tachyon Model,''}
\J{JHEP}{0104}{2001}{003}, {\tt hep-th/0102173};
{\sl ``Metamorphosis Of Tachyon Profile In Unstable D9-Branes,''}
\J{Phys.Rev.}{D65}{2002}{026006}, {\tt hep-th/0102174}.

\bibitem{Yi-con}
P.~Yi, 
{\sl ``Membranes from Five-Branes and Fundamental Strings from Dp
	Branes,''} 
\J{Nucl.Phys.}{B550}{1999}{214}, 
{\tt hep-th/9901159};\\
O.~Bergman, K.~Hori and P.~Yi,
{\sl ``Confinement on the Brane,''}
\J{Nucl.Phys.}{B580}{2000}{289}, 
{\tt hep-th/0002223};\\
M.~Kleban, A.~Lawrence and S.~Shenker, 
{\sl ``Closed strings from nothing,''}
\J{Phys.Rev.}{D64}{2001}{066002}, 
{\tt hep-th/0012081}.

\bibitem{GHoriY}
G.~Gibbons, K.~Hori and P.~Yi, 
{\sl ``String Fluid from Unstable D-branes,''}
\J{Nucl.Phys.}{B596}{2001}{136}, {\tt hep-th/0009061}.


\bibitem{KMM2}
D.~Kutasov, M.~Marino and G.~Moore, 
{\sl ``Remarks on Tachyon Condensation in Superstring Field Theory,''}
{\tt hep-th/0010108}.

\bibitem{KrLa}
P.~Kraus and F.~Larsen, 
{\sl ``Boundary String Field Theory of the DDbar System,''}
\J{Phys.Rev.}{D63}{(2001)}{106004}, {\tt hep-th/0012198}; \\
T.~Takayanagi, S.~Terashima and T.~Uesugi, 
{\sl ``Brane-Antibrane Action from Boundary String Field Theory,''}
\J{JHEP}{0103}{2001}{019}, {\tt hep-th/0012210}.

\bibitem{originalroll}
A.~Sen, 
{\sl ``Rolling Tachyon,''} 
\J{JHEP}{0204}{2002}{048}, 
{\tt hep-th/0203211}. 

\bibitem{ST}
S.~Sugimoto and S.~Terashima,
{\sl ``Tachyon Matter in Boundary String Field Theory,''}
\J{JHEP}{0207}{2002}{025}, {\tt hep-th/0205085}.

\bibitem{Minahan}
J.~A.~Minahan, 
{\sl ``Rolling the tachyon in super BSFT,''} 
\J{JHEP}{0207}{2002}{030}, 
{\tt hep-th/0205098}.

\bibitem{tacmat}
A.~Sen,
{\sl ``Tachyon Matter,''} 
\J{JHEP}{0207}{2002}{065}, 
{\tt hep-th/0203265}; 
{\sl ``Field theory of tachyon matter,''}
\J{Mod. Phys.Lett}{A17}{2002}{1797},
{\tt hep-th/0204143}.


\bibitem{Ue}
A.~Ishida and S.~Uehara, 
{\sl ``Gauge Fields on Tachyon Matter,''}
\J{Phys.Lett.}{B544}{2002}{353}, {\tt hep-th/0206102}.

\bibitem{UeTe}
S.~Terashima and T.~Uesugi,
{\sl ``On the supersymmetry of non-BPS D-brane,''}
\J{JHEP}{0105}{2001}{059},
{\tt hep-th/0104176}.

\bibitem{GHY}
G.~W.~Gibbons, K.~Hashimoto and P.~Yi, 
{\sl ``Tachyon Condensates, Carrollian Contraction of Lorentz Group, and
	Fundamental Strings,''} 
\J{JHEP}{0209}{2002}{061}, {\tt hep-th/0209034}.

\bibitem{ZS}
M.~Marino, 
{\sl ``On the BV formulation of boundary superstring field theory,''}
\J{JHEP}{0106}{2001}{059}, 
{\tt hep-th/0103089};\\
V.~Niarchos and N.~Prezas, 
{\sl ``Boundary Superstring Field Theory,''}
\J{Nucl.Phys.}{B619}{2001}{51}, {\tt hep-th/0103102}.

\bibitem{Tseytlin-mobius}
A.~A.~Tseytlin, 
{\sl ``Renormalization of Mobius Infinities and Partition Function
	Representation for the String Effective Action,''}
\J{Phys.Lett.}{B202}{1988}{81}.

\bibitem{Frolov}
S.~Frolov, 
{\sl ``On off-shell structure of open string sigma model,''}
\J{JHEP}{0108}{2001}{020}, {\tt hep-th/0104042}.

\bibitem{HT}
K.~Hashimoto and S.~Terashima, to appear.

\bibitem{HaKuMa}
J.~A.~Harvey, D.~Kutasov and E.~J.~Martinec,
``On the relevance of tachyons,''
{\tt hep-th/0003101}.

\bibitem{AsSuTe}
T.~Asakawa, S.~Sugimoto and S.~Terashima, 
{\sl ``Exact Description of D-branes via Tachyon Condensation,''}
\J{JHEP}{0302}{2003}{011}, {\tt hep-th/0212188}. 

\bibitem{bE}
P.~Mukhopadhyay and A.~Sen, 
{\sl ``Decay of Unstable D-branes with Electric Field,''}
\J{JHEP}{0211}{2002}{047}, {\tt hep-th/0208142};\\
S.~-J.~Rey and S.~Sugimoto, 
{\sl ``Rolling Tachyon with Electric and Magnetic Fields -- T-duality
	approach -----,''} 
\J{Phys.Rev.}{D67}{2003}{086008}, {\tt hep-th/0301049};
{\sl ``Rolling of Modulated Tachyon with Gauge Flux and Emergent
	Fundamental String,''} 
\J{Phys. Rev.}{D68}{2003}{026003}, {\tt hep-th/0303133};\\
C.~Kim, H.~B.~Kim, Y.~Kim and O-K.~Kwon, 
{\sl ``Electromagnetic String Fluid in Rolling Tachyon,''}
\J{JHEP}{0303}{2003}{008}, {\tt hep-th/0301076}.

\bibitem{tree}
A.~Sen, {\sl ``Open-Closed Duality at Tree Level,''}
\J{Phys.Rev.Lett.}{91}{2003}{181601}, {\tt hep-th/0306137}.

\bibitem{Yi}
H.-U.~Yee and P.~Yi, 
{\sl ``Open/Closed Duality, Unstable D-Branes, and Coarse-Grained Closed
	Strings,''}
{\tt hep-th/0402027}.

\bibitem{li-witten}
K.~Li and E.~Witten, 
{\sl ``Role of Short Distance Behavior in Off-Shell Open-String Field
	Theory,''} 
\J{Phys.Rev.}{D48}{1993}{853}, {\tt hep-th/9303067};\\
O.~Andreev,
{\sl ``Some Computations of Partition Functions and Tachyon Potentials
	in Background Independent Off-Shell String Theory,''} 
\J{Nucl.Phys.}{B598}{2001}{151}, {\tt hep-th/0010218}.

\bibitem{prop}
E.~S.~Fradkin and A.~A.~Tseytlin, {\sl ``Nonlinear Electrodynamics from
	Quantized Strings,''} \J{Phys.Lett.}{163B}{1985}{123};\\
A.~Abouelsaood, C.~G.~Callan, C.~R.~Nappi and S.~A.~Yost, 
{\sl ``Open Strings in Background Gauge Fields,''}
\J{Nucl.Phys.}{B280}{1987}{599}.

\bibitem{Herdeiro}
G.~W.~Gibbons and C.~A.~R.~Herdeiro, 
{\sl ``Born-Infeld Theory and STringy Causality,''}
\J{Phys.Rev.}{D63}{2001}{064006}, {\tt hep-th/0008052}.

\bibitem{HataTera}
H.~Hata and S.~Teraguchi, 
{\sl ``Test of the Absence of Kinetic Terms around the Tachyon Vacuum in
	Cubic String Field Theory,''} 
\J{JHEP}{0105}{2001}{045}, {\tt hep-th/0101162};\\
I.~Kishimoto and T.~Takahashi, 
{\sl ``Open String Field Theory around Universal Solutions,''}
\J{Prog.Theor.Phys.}{108}{2002}{591}, {\tt hep-th/0205275}.

\bibitem{TZ}
T.~Takahashi and S.~Zeze, 
{\sl ``Gauge Fixing and Scattering Amplitudes in String Field Theory
	Expanded around Universal Solutions,''} 
\J{Prog.Theor.Phys.}{110}{2003}{159}, 
{\tt hep-th/0304261}.

\bibitem{drukker}
N.~Drukker, 
{\sl ``Closed String Amplitudes from Gauge Fixed String Field Theory,''}
\J{Phys.Rev.}{D67}{2003}{126004}, {\tt hep-th/0207266};
{\sl ``On different actions for the vacuum of bosonic string field
	theory,''} 
\J{JHEP}{0308}{2003}{017}, {\tt hep-th/0301079}; \\
I.~Katsumata, T.~Takahashi and S.~Zeze, talk given at a JPS meeting; \\
Y.~Igarashi, K.~Ito, I.~Katsumata, T.~Takahashi and S.~Zeze, to appear.

\bibitem{stro}
A.~Strominger, 
{\sl ``Open String Creation by S-Branes,''}
{\tt hep-th/0209090}.

\bibitem{bl}
M.~Gutperle and A.~Strominger,
{\sl ``Timelike boundary Liouville theory,''}
\J{Phys.Rev.}{D67}{2003}{126002}, {\tt hep-th/0301038}; \\
F.~Larsen, A.~Naqvi and S.~Terashima,
{\sl ``Rolling tachyons and decaying branes,''}
\J{JHEP}{0302}{2003}{039}, {\tt hep-th/0212248}; \\
V.~Balasubramanian, E.~Keski-Vakkuri, P.~Kraus and A.~Naqvi,
{\sl ``String Scattering from Decaying Branes,''}
{\tt hep-th/0404039}.

\bibitem{fred}
S.~Fredenhagen and V.~Schomerus, 
{\sl ``On Minisuperspace Models of S-branes,''}
\J{JHEP}{0312}{2003}{003}, {\tt hep-th/0308205}.

\bibitem{Te}
S.~Terashima, 
{\sl ``A Construction of Commutative D-branes from Lower Dimensional
	Non-BPS D-branes,''} 
\J{JHEP}{0105}{2001}{059}, {\tt  hep-th/0101087}. 

\end{thebibliography}
\end{document}